\documentclass[aps,prl,preprint, showpacs]{revtex4}

\usepackage{graphicx}
\usepackage{dcolumn}
\usepackage{amsmath}

\begin{document}

\title{Spin-Hall conductivity of a spin-polarized two-dimensional electron
gas with Rashba spin-orbit interaction and
magnetic impurities}
\author{C. P.  Moca$^{1,2}$ and D. C. Marinescu$^3$}
\affiliation{
$^1$Department of Physics, University of Oradea, 410087 Oradea, Romania}
\affiliation{
$^2$Institute of Physics, Technical University Budapest, Budapest,
H-1521, Hungary}

\affiliation{$^3$Department of Physics and Astronomy, Clemson University, 29634,
Clemson}

\date{\today}
\begin{abstract}
The Kubo formula is used to calculate the spin-Hall conductivity
$\sigma_{sH}$ in a spin-polarized two-dimensional electron system
with Rashba-type spin-orbit interaction.  As in the case of the
unpolarized electron system, $\sigma_{sH}$ is entirely determined by
states at the Fermi level, a property that maintains in the presence
of magnetic impurities. In the clean limit, the spin-Hall
conductivity decreases monotonically with the Zeeman splitting, a
result of the ordering effect on the electron spins produced by the
magnetic field. In the presence of magnetic impurities, the
spin-dependent scattering determines a finite renormalization of the
static part of the fully dressed vertex correction of the velocity
operator that leads to an enhancement of $\sigma_{sH}$, an opposite
behavior to that registered in the presence of spin-independent
disorder. The variation of $\sigma_{sH}$ with the strength of the
Rashba coupling and the Zeeman splitting is studied.
\end{abstract}

 \maketitle
\section {Introduction}

Equivalent to a local, momentum-dependent effective magnetic-field,
the spin-orbit interaction (SOI) in two-dimensional (2D) electron
systems introduces a spin-dependent, chiral motion of the electrons
that is sensitive to the application of an electric field. This
property, that opens up the possibility of manipulating the electron
spins exclusively by electrical means, is at the root of the
tremendous amount of interest in understanding the electron dynamics
in the presence of SOI, given the potential applications to
spintronics.

One such example is the intrinsic spin-Hall effect, when a pure spin
current flows in a transverse direction under the action of an
electric field \cite{sinova, murakami}. The spin current is
polarized along the third perpendicular direction. The magnitude of
the spin-current response, described by the spin-Hall conductivity
$\sigma_{sH}$, reaches, in a clean system, a universal value
$e/8\pi$, independent of any sample parameters.

The behavior of the intrinsic spin-Hall effect \cite{sH} in the
presence of non-magnetic impurities has been a subject of intense
investigation. While analytic calculations led to a cancelation of
the effect even in the presence of infinitesimal impurity
concentration \cite{dimitrova}, numerical studies, done in finite
size samples \cite{murakami,marinescu, nikolic}, indicated that the
spin-Hall effect persists in mesoscopic samples, up to a certain
disorder strength. It has been shown that within the bulk, the
spin-Hall conductivity is decaying exponentially along a distance of
the order of magnitude of the spin precession length \cite{pascu}.
More recent reports indicate that the discontinuous variation of
$\sigma_{sH}$ in the infinite 2D system, from a finite value in the
clean system to zero in the presence of the infinitesimal disorder,
can be explained by introducing an additional dephasing effect
associated with the inelastic electron lifetime \cite{wang}. This
result suggests that the spin-Hall conductivity is enhanced by
interactions that introduce additional scattering of the the
electron spins and maintains a finite value even when the
clean-disordered transition is performed. Naturally, one wonders if
the opposite effect might be true. Are interactions leading to an
ordering of the spins, such as the Zeeman coupling to an external
magnetic field, acting as decreasing factors on $\sigma_{sH}$?

Inspired by these ideas, we proceed to a calculation of the spin
Hall conductivity in a 2D system with Rashba spin-orbit coupling,
spin-polarized by a static
magnetic field, perpendicular on the sample.  The alignment of the electron
spins along the direction of the magnetic field
counteracts the spatial disordering induced by the spin-orbit
coupling, leading in consequence to a diminished
contribution to the spin current.
Further, we consider magnetic impurities and study the competing
effects of the Zeeman splitting and spin-dependent impurity scattering. The
latter affects the magnitude of
$\sigma_{sH}$ through the
renormalization effects it induces on
the vertex corrections of the current operator.

The simple model we discuss below, that of a non-interacting 2D
spin-polarized electron gas with SOI and magnetic impurities, allows
the simultaneous investigation of the intrinsic anomalous Hall
effect, which would occur only when a finite magnetization is
present, and of the spin-Hall effect in the presence of a
distribution of magnetic scatterers, previously analyzed within a
paramagnetic system \cite{inoue}. Our calculation is based on the
Kubo formula, where we take into account the scattering of the
electrons on the magnetic impurities. The algorithm discussed here,
generalizes to spin transport the traditional treatment of the
off-diagonal anomalous Hall conductivity of Ref. \cite{dugaev}, a
method that has been also used with great success to investigate the
anomalous Hall effect in graphenes \cite{sinitsyn, sinitsyn_1}.
Within this framework we start by obtaining the impurity-averaged
single-electron Green's functions and the renormalized vertex
correction of the velocity operator. Then, we apply the Kubo formula
to estimate the spin-Hall conductivity. First, in the case of a
clean system, we use the exact eigenvalues-eigenstates of the
Hamiltonian and obtain an analytic result for $\sigma_{sH}$ which
show its dependence on the Zeeman splitting. Later, we use the
impurity averaged Green's functions and the vertex-corrected current
operator to estimate the spin-Hall conductivity in the presence of
magnetic impurities. Analytical expressions for $\sigma_{sH}$ are
derived as functions of the Zeeman splitting and the magnetic
impurity scattering.

\section{Theoretical Framework}

\subsection{Model Hamiltonian}

We consider a non-interacting two-dimensional (2D) electron gas with
Rashba-type spin-orbit coupling (proportional to the linear
momentum) in the presence of a magnetic field.  The system is
assumed to contain magnetic impurities. The magnetic field
$\mathbf{B}$, is oriented along the $\hat{z}$ direction and is
perpendicular on the layer. The resulting Zeeman splitting $E_Z =
2\gamma B$, proportional to the gyromagnetic factor $\gamma$ is
considered a parameter of the problem. The noninteracting,
single-particle Hamiltonian, written for an electron of wave-vector
$\mathbf{k}=\{k_x,k_y\}$ and kinetic energy $\epsilon_k =
\hbar^2k^2/2m$ in respect with the Fermi surface $\mu$, ($m$ is
approximated by the bare mass) is
\begin{equation}
H_0 = \epsilon_k + \alpha (k_y\sigma_x -k_x\sigma_y) - E_Z\sigma_z
\;, \label{eq:h1}
\end{equation}
where $\alpha$ designates the spin-orbit coupling constant, while
$\sigma_i,(i=x,y,z)$ are the Pauli matrices. In the two-dimensional
spin space, an elementary diagonalization procedure generates the two
eigenvalues
\begin{equation}
\mathcal{E}_{k,\pm} = \epsilon_k \mp \sqrt{\alpha^2k^2 + E_Z^2}
\label{eq:en}
\end{equation}
and the associated eigenstates of  the Hamiltonian:
\begin{equation}
\begin{array}{rr}
\psi_{+} = \left(\begin{array}{r}\cos \frac{\theta}{2}e^{i\phi/2}\\
- \sin \frac{\theta}{2}e^{-i\phi/2}\end{array}\right); \hspace{1cm}
&
\psi_{-} = \left(\begin{array}{r}\sin \frac{\theta}{2}e^{i\phi/2}\\
\cos \frac{\theta}{2}e^{-i\phi/2}\end{array}\right)
\end{array} \label{eq:vec}
\end{equation}
with $\cos \theta = E_Z/\Delta_k$, $\sin \theta =
\alpha\,k/\Delta_k$ and
\begin{equation}
\Delta_k = \sqrt{\alpha^2k^2 + E_Z^2}
\label{eq:delta}
\end{equation}
the effective Rashba gap.

When the magnetic impurities are present, an additional coupling
Hamiltonian has to be included in Eq.~(\ref{eq:h1}):
\begin{equation}
H_{imp} = J\,\mathbf{s}\cdot\mathbf{S}\;
\end{equation}
with $\mathbf{s}$ and $\mathbf{S}$ denoting the electron and the
impurity spin, respectively. The electron spin is treated like a
quantum mechanical observable, described in terms of the
spin-dependent creation and destruction operators at site $i$, $c_i
=(c_{i\uparrow},c_{i\downarrow} )$ and $c_i^\dag =
(c_{i\uparrow}^\dag,c_{i\downarrow}^\dag )$ by $\mathbf{s} =
\frac{\hbar}{2}c_i^\dag\, \boldsymbol{\sigma}\,c_i$. The impurity
spin is considered to be a classical variable, whose direction
$\mathbf{n}$, in spherical coordinates is specified by the angles
$\theta$ and $\phi$: $ \mathbf{S} =  S \mathbf{n}= S(\sin \theta
\cos \phi, \sin \theta \sin \phi, \cos \theta) $. In the spinor
representation, the coupling can be described by the matrix
\begin{equation}
U(\theta, \phi) = \left(\begin{array}{rr}\cos \theta&\sin \theta
e^{-i\phi}\\
\sin \theta e^{i\phi}&-\cos \theta\end{array}\right) =
\boldsymbol{\sigma}\cdot \mathbf{n} \label{eq:u}.
\end{equation}
With  $u= \hbar J\, S/2$, a rescaled exchange coupling, the
interaction Hamiltonian is then written as
\begin{equation}
H_{imp} = u\,\sum_ic_i^\dag\left( \boldsymbol{\sigma}\cdot \mathbf{n}\right)
c_i
\end{equation}
Throughout this analysis, the impurity scattering problem is treated
perturbatively, as we neglect the regime where the Kondo effect may
be important. In our approximation, the lifetime of the
quasiparticles at the Fermi level is evaluated for each band, as the
imaginary part of the self-energy in the second order perturbation
theory. At the same time, the shift of the chemical potential, due
to the real part of the self-energy, is not considered.

\subsection{Green's function, self-energy and current vertex correction}

\begin{figure}[h]
\centering
\includegraphics[width=2.5in]{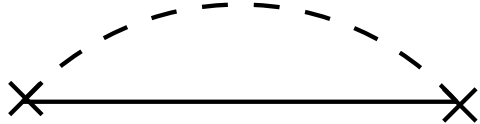}
\caption{Second order self-energy contribution due to magnetic impurities.
The solid
line represents the Green's function while the crosses describe the
interaction
with a single magnetic impurity. }
\label{fig:self_energy}
\end{figure}

The free electron Green's function is obtained from the single
particle Hamiltonian, Eq.~(\ref{eq:h1}) as a $2\times 2$ matrix in
the spin space:
\begin{equation}
G^0_k(\omega) = \frac{\omega-\epsilon_k + \mu + \alpha (k_y\sigma_x
- k_x\sigma_y) - E_Z\sigma_z}{\left[\omega-\mathcal{E}_{k,+}+\mu +
i\delta\, sgn\,(\omega)\right]\left[\omega-\mathcal{E}_{k,-}+\mu +
i\delta \,sgn (\omega)\right]} \label{eq:green1}
\end{equation}
with $\delta >0$ an infinitesimally small quantity. In the presence
of the impurities, $G^0_k(\omega)$ is modified to include the
effects of the elastic scattering. The relaxation time is given by
the imaginary part of the self-energy, which, in the lowest order
(see Fig. \ref{fig:self_energy}), is obtained from:
\begin{equation}
\Sigma (\omega) = n_i\,u^2\int \frac{d^2k}{(2\pi)^2}\int
\frac{d\Omega}{4\pi}U(\theta,\phi)G^0_k(\omega)U(\theta,\phi)
\end{equation}
As an explicit function of the effective Rashba gap and the Zeeman
splitting, the self-energy is given by:
\begin{eqnarray}
\Sigma_i(\omega) &= &\frac{n_i\,u^2}{2}\int \frac{d^2k}{(2\pi)^2}\int
\frac{d\Omega}{4\pi} \label{eq:self}
\left (\boldsymbol{\sigma}\cdot\mathbf{n}\right )  \nonumber\\
& \times &  \left . \left[ \frac{1}{\omega - \epsilon_k+\Delta_k+\mu
+i\delta{sgn}(\omega)}+ \frac{1}{\omega - \epsilon_k-\Delta_k+\mu
+i\delta {sgn}(\omega)}\right]
\left ( \boldsymbol{\sigma}\cdot \mathbf{n}\right )\right. \\
& + & \left (\boldsymbol{\sigma}\cdot \mathbf{n}\right )
\frac{E_Z\sigma_z}{\Delta_k}\left[\frac{1}{\omega - \epsilon_k
+ \Delta_k+\mu +i\delta {sgn}(\omega)} \right.  \left.   -\frac{1}{\omega -
\epsilon_k-\Delta_k +\mu +i\delta
{sgn}(\omega)}\right]
\left ( \boldsymbol{\sigma}\cdot \mathbf{n}\right )  \nonumber
\end{eqnarray}
Since the real part of Eq.~(\ref{eq:self}) just renormalizes the
Fermi energy, we focus only on its imaginary part, the one that
determines the quasiparticle lifetime at the Fermi level. In
contrast to the case of non-magnetic impurities, now the scattering
rates depend on the chirality of the band:
\begin{eqnarray}
\Im m \Sigma_i & =&  - \pi sgn(\omega)
\frac{n_i\, u^2}{2}\int\frac{d^2k}{(2\pi)^2}\int \frac{d\Omega}{4\pi} \\
& & \left(\boldsymbol{\sigma}\cdot\mathbf{n} \right )
\left[\delta(\omega-\epsilon_k+\Delta_k+\mu)+\delta(\omega-\epsilon_k-\Delta
_k+\mu)\right]
\left (\boldsymbol{\sigma}\cdot \mathbf{n}\right) \nonumber \\
& +&  \left (\boldsymbol{\sigma}\cdot \mathbf{n}\right )
\frac{E_Z\sigma_z}{\Delta_k}
\left[\delta(\omega-\epsilon_k+\Delta_k+\mu)-\delta(\omega-\epsilon_k-\Delta
_k+\mu)\right]
\left (\boldsymbol{\sigma}\cdot\mathbf{n}\right)\nonumber
\end{eqnarray}
The momentum space integral is processed by changing to an integral
over energy, $\int d^2k/(2\pi)^2 \rightarrow \int d\epsilon
N_0(\epsilon)$ where $N_0(\epsilon) = m/2\pi$ the density of
states at the Fermi surface. After performing the integrals over the solid
angles,
and  some standard manipulations, we finally obtain:
\begin{equation}
\Im m \Sigma_i(\omega) = -\frac{\pi n_i u^2}{2}\mbox{sgn}(\omega)
\frac{m}{2\pi}\left[\frac{1}{\left|1-\frac{m\alpha^2}{\Delta_{k_{F+}}}\right
|}\\
+\theta(\mu-E_Z)\frac{1}{\left|1-\frac{m\alpha^2}{\Delta_{k_{F-}}}\right|}
\right]
\end{equation}
We recognize that
\begin{equation}
\begin{array}{rr}
N_+  =
\frac{m}{2\pi}\left|1-\frac{m\alpha^2}{\Delta_{k_{F+}}}\right|^{-1}
; \hspace{1cm} & N_-  =  \theta(\mu -
E_Z)\frac{m}{2\pi}\left|1-\frac{m\alpha^2}{\Delta_{k_{F-}}}\right|^{-1}
\end{array}
\end{equation}
are the densities of states in the chiral bands, allowing us to
define the symmetric and antisymmetric scattering rates:
\begin{equation}
\begin{array}{rr}
\frac{1}{\tau} = \pi n_i u^2(N_+ + N_-); \hspace{1cm} &
\frac{1}{\bar{\tau}}= -\frac{1}{3}\pi n_i
u^2 E_Z\left(\frac{N_+}{\Delta_+}-\frac{N_-}{\Delta_-}\right)
\end{array}
\end{equation}
where Eq.~(\ref{eq:delta}) was employed. Thus,
\begin{equation}
\Im m \Sigma_i (\omega) = -\frac{1}{2}\left(\frac{1}{\tau} +
\frac{1}{\bar{\tau}}\sigma_z\right)\mbox{sgn}(\omega)
\end{equation}
We introduce the band-dependent impurity scattering times,
$\tau_\pm^{-1} = \tau^{-1}\pm\bar{\tau}^{-1}$ and define $\Gamma =
1/2\tau$, $\bar{\Gamma} = 1/2\bar{\tau}$,  $\Gamma_{\pm} =
1/2\tau_{\pm}$ . With these notations, the impurity-averaged Green's
function $G_k(\epsilon)$  becomes
\begin{equation}
G_k(\omega) = \frac{\omega + i\Gamma\, \mbox{sgn}(\omega)- \epsilon_k + \mu
+ \alpha
(k_y\sigma_x - k_x\sigma_y)-[E_Z+
i\bar{\Gamma}\,
\mbox{sgn}(\omega)]\sigma_z}{\left[\omega-\mathcal{E}_{k,+}+\mu +
i\Gamma_+\,\mbox{sgn}(\omega)\right]\left[\omega-\mathcal{E}_{k,-}+\mu +
i\Gamma_{-}\,\mbox{sgn}(\omega)\right]}\label{eq:gf}
\end{equation}
Eq.~(\ref{eq:gf}) generates the retarded (R) and advanced (A)
Green's functions, $G_{k}^{(A,R)}(\omega)$, in the second-order
perturbation theory.

The next ingredient needed for computing the spin-Hall conductivity
is the current vertex, involved in the calculation of the
polarization bubble when multiple scattering events on the magnetic
impurities are considered.
\begin{figure}[h]
\centering
\includegraphics[width=4.0in]{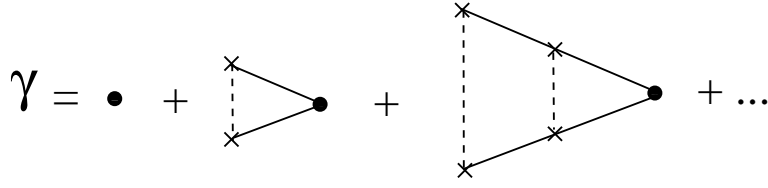}
\caption{Diagrammatic representation of the  current vertex. The dots
represents the bare  current, the upper/lower solid lines are the
retarded/advanced
Green's function and the crosses  represent the interaction with
the magnetic impurities.}
\label{fig:vertex}
\end{figure}
We determine the vertex-corrected current as the ladder series
expressed in Fig. \ref{fig:vertex}, whose equivalent analytical
equation is:
\begin{equation}
\gamma_x (\epsilon,\omega) = v_x +
n_i\, u^2\int\frac{d\Omega}{4\pi}\int
\frac{d^2k}{(2\pi)^2}\left(\boldsymbol{\sigma}\cdot\mathbf{n} \right )
G^R_k(\epsilon)\,\gamma_x^2\,(\epsilon,\omega)G_k^A(\epsilon+\omega)\left(
\boldsymbol{\sigma}\cdot\mathbf{n} \right )
\label{eq:vertex_green}
\end{equation}
In the static limit, when $\epsilon \rightarrow 0$ and $\omega
\rightarrow 0$, we write
\begin{equation}
\gamma_x =
\frac{k_x}{2m}-\alpha \sigma_y
+n_iu^2\int\frac{d\Omega}{4\pi}(\boldsymbol{\sigma}\cdot
\mathbf{n})G^R_k(0)\,\gamma_x\,G_k^A(0) (\boldsymbol{\sigma}\cdot
\mathbf{n}) \label{eq:vertex}
\end{equation}
where the advanced (A) and retarded (R) electron Green's functions
are obtained from the static limit of  Eq. (\ref{eq:gf}). A solution
to Eq.~(\ref{eq:vertex}) can be obtained in the form of an
expansion: $ \gamma_x = k_x/m -\gamma_x^{\mu} \sigma_{\mu}$. First,
a simple analysis shows that two components of the static part of
the dressed vertex vanish: $\gamma_x^{x} = \gamma_x^z= 0$. The
remaining, non-zero component of the vertex function is $\gamma_x^y$
expressed as:
\begin{equation}
\gamma_x^y =\alpha \left (1+ \frac{2 n_i\,
u^2}{3}\int\frac{d^2k}{(2\pi)^2}\frac{\epsilon_k(\epsilon_k -
\mu)}{A_+A_-}\right )
\left ({1+\frac{n_i\, u^2}{3}\int\frac{d^2k}{(2\pi)^2}
\frac{(\epsilon_k -\mu)^2-E_Z^2}{A_+A_-}}\right )^{-1}
\end{equation}
where
\begin{equation}
A_\pm = \left(\mu - \mathcal{E}_{k\pm}\right
)^2 + \Gamma_{\pm}^2. \label{eq:apm}
\end{equation}
In the case of non-magnetic impurities the vertex coefficient
$\gamma_x^y$ cancels when the Fermi level is in the upper band (both
bands are occupied), leading to the disappearance of the spin-Hall
effect in the thermodynamic limit when any amount of disorder is
present. This is not the case, however, when magnetic impurity
scattering occurs, since now the static part of the dressed vertex
is larger than the bare Rashba coupling for any ratio $\mu/E_Z> -1$.
In our model the Rashba coupling is the static part of the bare
vertex.

\begin{figure}[t]
\centering
\includegraphics[width=4.0in]{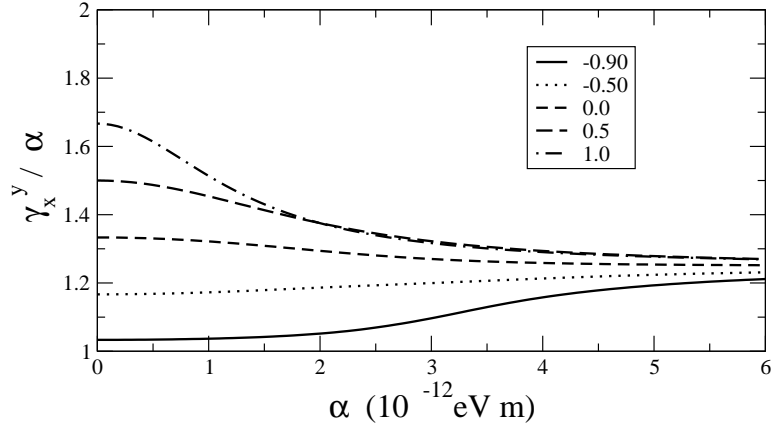}
\caption{$\gamma_x^y$ as function of the Rashba coupling strength
for different ratios of chemical potential versus the magnetic
energy $\mu/E_Z$. Here $E_Z = 1.1 \times 10^{-4} eV$ and the
effective electron mass is the bare one $m_e$.} \label{fig:gama_x_y}
\end{figure}

As can be seen in Fig. \ref{fig:gama_x_y}, in the extreme case when
the chemical potential satisfies $\mu = E_Z+\delta_E$ (with
$\delta_E$ some positive infinitesimal energy), so just one band is
partially occupied,  the vertex is practically not renormalized and
takes the bare value $\alpha$. The renormalization is more
pronounced as the band is gradually filled. The largest
renormalization is obtained when both bands are occupied $\mu >
E_Z+\delta_E$ . Increasing the ratio $\mu/E_Z$ above $1$ does not
lead to a larger renormalization. Therefore, for any filling factor,
in the thermodynamic limit, the spin-Hall conductivity is finite
when magnetic impurities are present in the system, irrespective of
how strong/weak the interaction potential is. For the experimentally
accessible values of the Rashba coupling strength of $5-6\times
10^{-12} eV m$ \cite{nitta} we present the behavior of the vertex
function in Fig. \ref{fig:gama_x_y}.
An external magnetic field of $B=1T$ is considered to be applied, so
that the Zeeman energy is approximately $E_Z = 1.1 \times 10^{-4}
eV$.

\section{Spin-Hall conductivity}
In this section we present analytical results for the spin-Hall
conductivity. First we compute $\sigma_{sH}$ for a non-interacting
system. Because in the clean limit the eigenvectors and
eigenenergies are exactly known, the Kubo formula can be employed,
and an analytical result is obtained in agreement with previous
analytical work.  When magnetic impurity effects are investigated,
Kubo formalism in term of exact eigenstates/eigenenergies is no
longer suited and the causal Green's function method has to be
considered. For that, the impurity averaged Green's functions and
the vertex correction derived in the previous section are needed.

\subsection{Spin Hall conductivity for the non-interacting system. Exact
result}

 The Kubo
formula that determines the spin-Hall conductivity in the clean
system is written in the chiral basis of states, Eq.~(\ref{eq:vec}):
\begin{equation}
\sigma_{sH}= e\hbar \sum_{n\neq n'} \int
\frac{d^2\mathbf{k}}{(2\pi)^2}\left(f_{kn'}-f_{kn}\right) \Im m
\frac{<k\,n'|j_x^z|k\,n><k\,n|v_y|k\,n'>}{(\mathcal{E}_{kn}-\mathcal{E}_{kn'
})^2} \label{eq:kubo}
\end{equation}
where $n$ is a band index, in our case $n, n' = \pm$. The electron
velocity along the $y$ direction is $ v_y = k_y/m + \alpha \sigma_x$
and the $\hat{z}$-polarized current propagating in the $x$ direction
is $j_x^z = \hbar/4\{v_x, \sigma_z\} = \hbar\,k_x/2m \,\sigma_z$.
Upon the insertion of their matrix elements, evaluated in the chiral
basis, in Eq.~(\ref{eq:kubo}) we obtain:
\begin{equation}
\sigma_{xy}^z = \frac{\alpha^2 e}{16\pi m}\int \frac{k^3
dk}{\Delta_k^3}(f_{k+}-f_{k-})
\end{equation}
with $f_{k\pm}$ the Fermi-Dirac distributions corresponding to the
two bands. In the absence of the magnetic field, when
$E_Z\rightarrow 0$, we recover the well known result \cite{sinova}
$\sigma_{xy}^z = e/ 8\pi$ for a clean two dimensional electronic
system. For a finite Zeeman splitting, an analytical result can be
derived in terms of the Fermi energies of the chiral bands:
\begin{equation}
\sigma_{xy}^z = \frac{e}{8\pi}\frac{1}{m\alpha^2}\left[\frac{E_Z^2+
m\alpha^2\epsilon_{F+}}{\sqrt{E_Z^2+2m\alpha^2\epsilon_{F+}}}
-\frac{E_Z^2+
m\alpha^2\epsilon_{F-}}{\sqrt{E_Z^2+2m\alpha^2\epsilon_{F-}}}\right]
\label{eq:sigma}
\end{equation}
a result that shows that even the simple presence of an external
magnetic field leads to a non-universal value for the spin-Hall
conductivity.

\subsection{Spin-Hall conductivity using the causal Green's function. The
role of disorder}
In this section we derive an analytical expression for the
spin-Hall conductivity when both magnetic impurities and
external magnetic field are considered.
The Kubo formula for the spin-Hall conductivity written for the
impurity averaged Green's function gives:
\begin{eqnarray}
\sigma_{xy}^z (\omega)  & = & \frac{e}{\omega}\,
Tr\int\frac{d\epsilon}{2\pi}\int
\frac{d^2k}{(2\pi)^2}\left<j_y^z f(\epsilon)\, (G^R(\epsilon)-G^A(\epsilon)
)\gamma_x G^A(\epsilon -\omega) \right . \nonumber \\
& &\left . -j_y^z G^R (\epsilon) \gamma_x  f(\epsilon -\omega)(G^R (\epsilon
-\omega)-G^A (\epsilon -\omega))\right>
\label{eq:kubo_green}
\end{eqnarray}
where the Green's functions include the scattering lifetimes and
$<\ldots>$ represents the impurity configuration average. There are
two types of contributions to the integral,
Eq.~(\ref{eq:kubo_green}): from states below the Fermi level and
from states close to the Fermi level \cite{yang}. The contribution
from states well below the Fermi level can be neglected in the limit
$\alpha p_F \ll \epsilon_F$, or $\epsilon_F \tau_{\pm} \gg 1$,
because it contains only combinations of the form $G^R G^R$ and $G^A
G^A$ \cite{schwab}. At the same time, magnetic impurities, have
practically no effects on these states due to small scattering
rates, as compared to their energies. In stark contrast, states
close to the Fermi energy are strongly effected once their energy
becomes comparable to the scattering time $1/\tau_{\pm}$.

\begin{figure}[tb]
\centering
\includegraphics[width=4.0in]{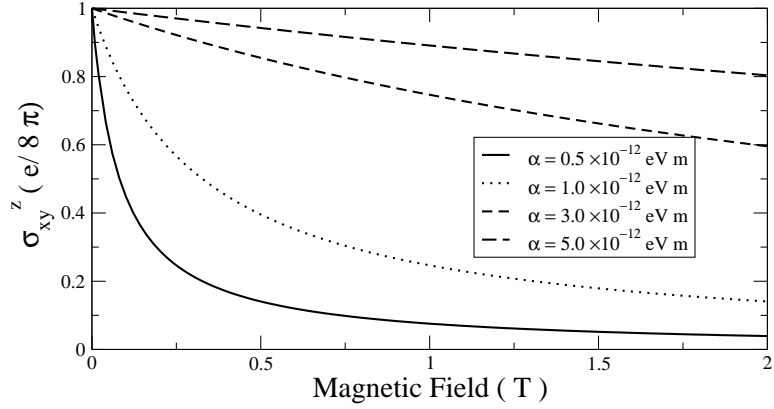}
\caption{Magnetic field dependence of the spin-Hall conductivity for
different SOI couplings when no vertex correction is considered
($\gamma_x^y \rightarrow \alpha$). For the case of magnetic
impurities this is a good approximation. Here $\mu /E_Z =1.5$}
\label{fig:magnetic_field}
\end{figure}
\begin{figure}[b]
\centering
\includegraphics[width=4.0in]{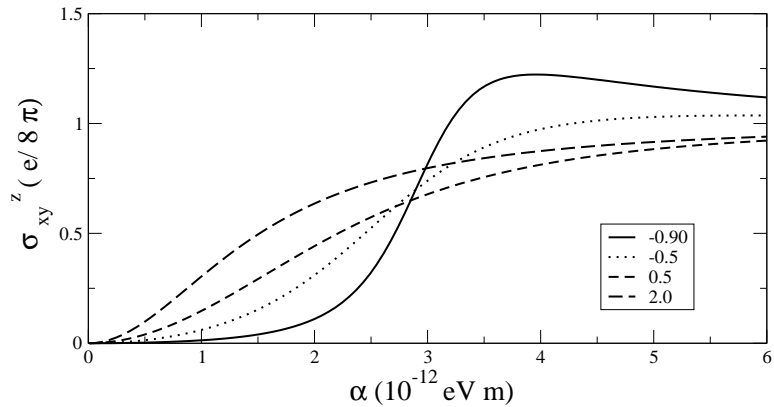}
\caption{Spin-Hall conductivity as function of the Rashba spin-orbit
interaction strength
for different ratios $\mu /E_Z$. Here an external field of $B=1T$ is
considered.}
\label{fig:rashba}
\end{figure}

The vertex correction becomes important when processes across the
Fermi surface are considered, so that one $G^{(R)}$ and one
$G^{(A)}$ enters the vertex equation (\ref{eq:vertex}). By taking
the zero frequency limit in  Eq. (\ref{eq:kubo_green}) the spin-Hall
conductivity becomes:
\begin{equation}
\sigma_{xy}^z = \frac{e}{8 \pi}  \,4 \alpha \gamma_x^y
\int\frac{d^2k}{(2\pi)^2}\frac{\epsilon_k}{A_-A_+}
\label{eq:sigma2}
\end{equation}
with $A_\pm$ given by Eq.~(\ref{eq:apm}).  Simple algebraic
manipulation gives an exact result for the spin-Hall conductivity in
terms of the density of states for the chiral bands.

\begin{equation}
\sigma_{xy}^z = \frac{e}{8 \pi}  \,4\pi \alpha \gamma_x^y \Gamma
\frac{1}{(\mathcal{E}_1 -\mathcal{E}_2)^2}\left (
N_0(\mathcal{E}_1)\frac{\mathcal{E}_1}{\Gamma_1}
+N_0(\mathcal{E}_2)\frac{\mathcal{E}_2}{\Gamma_2} \right )
\label{eq:sigma3}
\end{equation}
where we have introduced the quantities:
$\mathcal{E}_{1,2} = \mu+m\alpha^2 \pm (m^2 \alpha^4 +2 \mu m \alpha^2+
E_Z^2)^{1/2} $
and
\begin{equation}
\Gamma_{1,2} = \mp \frac{\mu -\mathcal{E}_{1,2}-(E_Z^2+2 m
\alpha^2\mathcal{E}_{1,2}) }{2\tau_+(\mathcal{E}_1 - \mathcal{E}_2)}
\mp \frac{\mu -\mathcal{E}_{1,2}+(E_Z^2+2 m \alpha^2\mathcal{E}_{1,2})
}{2\tau_-(\mathcal{E}_1 - \mathcal{E}_2)}
\end{equation}

In the absence of magnetization ($E_Z \rightarrow 0$), when no
vertex corrections are considered ($\gamma_x^y \rightarrow \alpha$),
and in the weak disorder limit,  Eq.~(\ref{eq:sigma2})  generates
the well-known universal expression of the spin Hall conductivity,
$\sigma_{xy}^z  = e/8\pi$. One important observation is that, in
contrast to the case of unpolarized disorder, when the static
component of the dressed velocity is renormalized to zero by the
vertex correction, here the vertex corrections lead to an
enhancement of $20-50\% $ of the bare static velocity (see Fig.
\ref{fig:gama_x_y}). This observations shows that, in the case of
magnetic impurities, even when  vertex correction for the velocity
are neglected, a good enough approximation for the spin-Hall
conductivity is obtained. In this limit ($\gamma_x^y \rightarrow
\alpha$) we present in Fig. \ref{fig:magnetic_field} and Fig.
\ref{fig:rashba} typical behaviors for the spin-Hall conductivity as
function of the external field as well as function of the spin-orbit
strength (similar curves are obtained also when the vertex is
considered). Typically, for a given $\alpha$, the magnetic field
reduces the strength of the spin-Hall effect. This can be, in
principle understood, by considering the different polarization
effects of the magnetic field, that statically orientates the
electron spins from both chiral bands along its direction, and the
Rashba interaction that induces an in-plane, dynamic polarization.
The larger the ratio $m\alpha^2/E_Z$ is, the stronger the spin-Hall
effect is, and in limit of zero magnetic field the universal
expression for the spin-Hall conductivity is reobtained.

\section{Conclusions}

The present work addresses an important topic in the field of
spin-Hall effect, that is the effect of magnetic impurities and the
role of a Zeeman term on the spin-Hall conductivity. We have
obtained simple but robust analytical results for the spin-Hall
conductivity, which in some particular limits converge to the
previous known results \cite{sinova, sH, inoue}.

First we find that the spin-Hall conductivity is no longer universal
in the presence of a magnetic field even in the clean limit.
The most important observation is related to the behavior
of the system when magnetic impurities are present. In this case
vertex correction leads to an enhance of the spin-Hall effect,
contrary to the case of non-magnetic impurities where the
static part of the fully dressed
vertex identically vanishes in the weak scattering limit. This allows
us to conclude that the bare vertex is a good approximation when
computing the spin-Hall conductivity, and that the bare bubble diagram
is good enough when computing the spin-Hall conductivity in the
presence of magnetic impurities.

\section*{Acknowledgments}
One of us (CPM) gratefully acknowledges support from the Romanian
Science Foundation,  grants CNCSIS/2006/1/97 and CNCSIS/2007/1/780 and
by the Hungarian Grants OTKA Nos. NF061726 and T046303.




\begin{thebibliography}{99}
\bibitem{sinova} Sinova J, Culcer D, Niu Q, Sinitsyn N A, Jungwirth T and
MacDonald A H 2004 \emph{Phys. Rev. Lett. }{\bf 92} 126603


\bibitem{murakami}  Sugimoto N, Onoda S, Murakami S and Nagaosa N 2006
\emph{Phys. Rev.} B {\bf 73} 113305


\bibitem{sH} Schliemann J and Loss D 2004 \emph{Phys. Rev.} B {\bf 69}
165315 \\
Bernevig B A, Hu J, Mukamel E, and Zhang S C 2004 \emph{Phys. Rev.} B {\bf
70} 113301\\
Bernevig B A and  Zhang S C  2005 \emph{Phys. Rev.} B {\bf 72} 115204\\
Rashba E I, 2003 \emph{Phys. Rev.} B {\bf 68} 241315
Burkov A A, Nunez  A S and  MacDonald A H,  2004 \emph{Phys. Rev.} B {\bf
70} 155308


\bibitem{dimitrova} Dimitrova O 2005 \emph{ Phys. Rev.} B {\bf 71} 245327


\bibitem{marinescu}  Moca C P and  Marinescu D C 2005 \emph{Phys. Rev.} B
{\bf 72} 165335


\bibitem{nikolic}  Nikoli\' c B K,  Souma S,  Z\^ arbo L  and  Sinova J 2005
\emph{Phys. Rev. Lett.} {\bf 95} 046601


\bibitem{pascu} Moca C P and Marinescu D C 2007 \emph{Phys. Rev.} B {\bf 75}
035325


\bibitem{wang} Wang P and Li, You-Quan, \emph{Preprint} cond-mat/0701425


\bibitem{inoue} Inoue J, Kato T, Ishikawa Y, Itoh H, Bauer G and Molenkamp L
W 2006 \emph{Phys. Rev. Lett.} {\bf 97} 046604


\bibitem{dugaev} Dugaev V K, Bruno P, Taillefumier M, Canals B and Lacroix C
2005 \emph{ Phys. Rev. }B { \bf 71} 224423


\bibitem{sinitsyn} Sinitsyn N A, MacDonald A H, Jungwirth T Dugaev V K and
Sinova J 2007 \emph {Phys. Rev.} B {\bf 75} 045315


\bibitem{sinitsyn_1} Sinitsyn N A,  Hill J E,  Min H, Sinova J and MacDonald
A H 2006 \emph {Phys. Rev. Lett.} {\bf 97} 106807


\bibitem{nitta} Nitta J, Akazaki T, Takayanagi H and Enoki T 1997
\emph{Phys. Rev. Lett.} {\bf 78} 1335


\bibitem{yang} Yang M F and Chang M C 2006 \emph{Phys. Rev. B} {\bf 73}
073304


\bibitem{schwab} Raimondi R and Schwab P 2005 Phys. Rev. B {\bf 71} 033311




\end{thebibliography}
\end{document}